\documentclass{PoS}
\usepackage{amsmath}
\usepackage{comment}

\title{Multi GPU Performance of Conjugate Gradient Solver with
  Staggered Fermions in Mixed Precision}

\ShortTitle{Mixed Precision Multi GPU Conjugate Gradient Solver}

\author{\speaker{Yong-Chull Jang}, Hyung-Jin Kim, and Weonjong Lee \\
  Lattice Gauge Theory Research Center, FPRD, and CTP \\
  Department of Physics and Astronomy,
  Seoul National University, Seoul, 151-747, South Korea \\
  E-mail: \email{wlee@snu.ac.kr}}

\abstract{%
  GPU has a significantly higher performance in single-precision
  computing than that of double precision.
  Hence, it is important to take a maximal advantage of the single
  precision in the CG inverter, using the mixed precision method.
  We have implemented mixed precision algorithm to our multi GPU
  conjugate gradient solver.
  The single precision calculation use half of the memory that is used
  by the double precision calculation, which allows twice faster data
  transfer in memory I/O.
  In addition, the speed of floating point calculations is 8 times 
  faster in single precision than in double precision.
  The overall performance of our CUDA code for CG is 145 giga flops per GPU
  (GTX480),  which does not include the infiniband network communication. 
  If we include the infiniband communication, the overall performance is
  36 giga flops per GPU (GTX480).
}

\FullConference{ The XXIX International Symposium on Lattice Field Theory - Lattice 2011\\
July 10-16, 2011\\
Squaw Valley, Lake Tahoe, California}

\begin{document}

\def\Tr{\mbox{Tr}}

\def\a{\alpha}
\def\b{\beta}
\def\c{\chi}
\def\d{\delta}
\def\e{\epsilon}              
\def\f{\phi}                  
\def\g{\gamma}
\def\h{\eta}
\def\i{\iota}
\def\j{\psi}
\def\k{\kappa}
\def\l{\lambda}
\def\m{\mu}
\def\n{\nu}
\def\o{\omega}
\def\p{\pi}               
\def\th{\theta}                
\def\r{\rho}                 
\def\s{\sigma}               
\def\t{\tau}
\def\u{\upsilon}
\def\x{\xi}
\def\z{\zeta}
\def\D{\Delta}
\def\F{\Phi}
\def\G{\Gamma}
\def\J{\Psi}
\def\L{\Lambda}
\def\O{\Omega}
\def\P{\Pi}
\def\Q{\Theta}
\def\S{\Sigma}
\def\U{\Upsilon}
\def\X{\Xi}

\section{Introduction}
CPU has been improving its computing performance but does not yet
quench the thirst of those demanding users who need more computing
power for their numerical challenges such as lattice QCD.
Graphic processing units (GPU) opens a new era for high performance
computing.
GPU is originally designed to handle 3-dimensional graphic images.
To achieve extremely high performance with geometric data, GPU is
designed of simple and tiny processors.
More modules are used for the data processing and, 
not for the data cache, nor for the flow control.
Hence, GPU are very different from typical CPUs by construction.
GPUs are very appropriate for highly intensive and parallelized
scientific computation.
At the beginning, programming GPU was quite challenging and difficult. 
Recently, Nvidia has introduced the CUDA library, which allow the users
to program the code for GPU easily. 

Since then, there have been several ways to program the GPU code:
the Nvidia CUDA, Open Graphic Library (Open GL), 
and Open Computing Language (Open CL) APIs.
In this paper, we focus on CUDA and its applications.
The CUDA provides us a user-friendly programming environment based on
the C, C++ programming language for GPU.
All of our CPS library codes are compiled and tested in CUDA version
3.2 and compute capability 1.3 mode.
We made the CUDA version of CG subroutines that were implemented as a
part of the Columbia Physics System (CPS) library.

\section{Conjugate gradient method}
Conjugate gradient (CG) algorithm \cite{john:1971} is an iterative
method for solving a linear algebraic equation of the following form.
\begin{equation}
\mathbf{b} = A\mathbf{x}\,,
\end{equation}
where $A$ is a $n \times n$ positive definite Hermitian matrix. 
$\mathbf{x}$ and $\mathbf{b}$ are complex vectors in the n dimensional
space.
Matrix $A$ and vector $\mathbf{b}$ are given and $\mathbf{x}$ is
a solution vector that we want to obtain.
Using the CG method we can get the solution $\mathbf{x}$ up to the
numerical precision what we want to achieve.
\begin{figure}[h]
\begin{center}
\framebox{\parbox{10cm}{\begin{quote}
$\mathbf{r} = \mathbf{b} - A\mathbf{x} \mbox{\qquad\qquad\quad\phantom{1}} \mathbf{r} \mbox{: residual vector}$\\
$\mathbf{d} = \mathbf{r} \mbox{\qquad\qquad\qquad\qquad} \mathbf{d} \mbox{: directional vector}$\\
$\delta_{new} = \mathbf{r}^{\dagger}\mathbf{r}\mbox{\qquad\qquad\qquad} \epsilon \mbox{: tolerance}$\\
$\delta_{0} = \delta_{new}$\\
for$(i=0; i<N_{dim} \& \delta_{new} > \epsilon^2\delta_{0}; \mbox{++}i)\{$\\
\mbox{\quad}$\alpha = \delta_{new}/\mathbf{d}^{\dagger}A\mathbf{d}$\\
\mbox{\quad}$\mathbf{x} = \mathbf{x} + \alpha \mathbf{d}$, 
\mbox{\quad}$\mathbf{r} = \mathbf{r} - \alpha A\mathbf{d}$\\
\mbox{\quad}$\delta_{old} = \delta_{new}$, 
\mbox{\quad}$\delta_{new} = \mathbf{r}^{\dagger}\mathbf{r}$ \\
\mbox{\quad}$\beta = \delta_{new}/\delta_{old}$, 
\mbox{\quad}$\mathbf{d} = \mathbf{r} + \beta \mathbf{d}$\\
$\}$
\end{quote}}}
\end{center}
\caption{Conjugate gradient
  algorithm\label{CG_structure}}
\end{figure}
In Fig.~\ref{CG_structure}, we show the structure of CG algorithm.
In the CG sequence, we have a number of linear algebra equations
such as vector addition, dot product, and scalar multiplication
and so on.
All of these linear algebra operations are implemented using CUDA
library.
Because most of these operations are not dominant part in CG
operation, there is no special applied optimization for those
functions except Dirac operation.
In Fig.~\ref{CG_structure}, $A\mathbf{d}$ and $A\mathbf{x}$ are Dirac
operations with staggered fermions \cite{PhysRevD.16.3031}. Basically, the
Dirac operation is a matrix-vector multiplication.
This is the most dominant part in CG sequence.
The matrix $A$ is defined as follows.
\begin{equation}
\mathbf{h} = A\mathbf{\chi}
\end{equation}
\begin{equation}
A = -D^{2} + m^{2} \quad \mbox{(m is quark mass)}
\end{equation}
\begin{equation}
D_{x,y} = U_{\mu}(x)\delta_{y,x+\mu} 
- U^{\dagger}_{\mu}(x-\mu)\delta_{y,x-\mu}
\end{equation}
\begin{equation}
D\chi(x) = \sum_{\mu}U_{\mu}(x)\chi(x+\mu) 
- U^{\dagger}_{\mu}(x-\mu)\chi(x-\mu)
\end{equation}
Here, the phase factor $\eta_\mu(x)$ is multiplied in advance to the
gauge link $U_\mu$ at the gauge link preconditioning part of CPS
library.
$h$ is a given as a source vector and $\chi$ is a staggered fermion
field which corresponds to the CG solution.
At one site of the lattice, a single Dirac operation $D\chi(x)$ needs
1584 bytes of data transfer and 576 number of floating point
calculations.
Let us consider a MILC fine lattice of $28^3 \times 96$.
A single Dirac operation $D\chi(x)$ over the entire, even sites of the
lattice needs 0.6 billion number of floating point calculations.
And it also needs 1.6 giga bytes of data transfer.
As a result, when we use GPUs for CG operations, it is easy to find
out that the bottle neck is on the data transfer rather then numerical
operation.

\section{Mixed Precision CG and implementation on CUDA}
\subsection{Mixed Precision CG}
Historically, there exists a significant gap between single precision
performance and double precision performance in GPUs.
In the case of the Nvidia GTX480 GPU, the single precision calculation
runs 8 times faster than the double precision calculation.

How much we can accelerate the program is depend on the ratio of
arithmetic operations and data I/O\footnote{I/O means input and
  output to the GPU memory.}.
The data I/O access is dominant in CG.
It is the main bottle-neck in CG algorithm.  Here, we do not include
the infiniband network communication in the data I/O.
So, by using the single precision data, the main benefit from it is
that we can reduce the data traffic by a factor of 2, between GPU
processor and GPU device memory.

To improve this performance of the CG program, the mixed precision method
has been used.
Mixed precision is implemented by iterative refinement
algorithm\cite{springerlink:10.1007/BF02162558}.
The main idea of the iterative refinement is using two types of iterative
loops to get the true solution value.
At first, by using the single precision iteration, we can approach
fast to the roughly estimated solution within inner loop tolerance.
And next, double precision or more precise iteration can be used to
get the more accurate solution within the outer loop error range.
The iterative refinement procedure in CG is illustrated in
Fig.~\ref{Mixed_CG}.
\begin{figure}[h]
\begin{center}
\framebox{\parbox{10cm}{\begin{quote}
$\mathbf{r_0} = \mathbf{b} - A\mathbf{x_0} \mbox{\qquad\qquad\quad} \mathbf{r} \mbox{: residual vector}$\\
while$(\|\mathbf{r_k}\| > \epsilon\|\mathbf{r_0}\|)\{$\\
\mbox{\quad}$\{$\\
\mbox{\qquad}$-$ \mbox{Inner Loop} $-$ \\
\mbox{\qquad}Solve\mbox{\quad}$A \mathbf{y} = \mathbf{r_k} \mbox{\quad within\quad}\epsilon^{in} $\\
\mbox{\quad}$\}$\\
\mbox{\quad}$\mathbf{x_{k+1}} = \mathbf{x_k} + \mathbf{y}$\\
\mbox{\quad}$\mathbf{r_{k+1}} = \mathbf{b} - A\mathbf{x_{k+1}}$\\
\mbox{\quad}$\mbox{k = k+1}$\\
$\}$
\end{quote}}}
\end{center}
\caption{Mixed precision CG algorithm($\epsilon^{in}$ is tolerance for
  low precision inner loop, $\epsilon$ means global convergence for
  overall CG sequence)\label{Mixed_CG}}
\end{figure}
In the inner loop, $A\mathbf{y} = \mathbf{r_k}$ is solved by
using single precision CG iteration which gives a approximate
solution to the correction terms to the outer loop solution.
From this method, we can replace most (\textbf{99.8\%}) of slow double
precision iterations by fast single precision iterations, while
preserving the total number of CG iterations.
As a result, the CG algorithm can run about three times faster.

\subsection{Corrected mixed precision CG}
It is possible to improve the performance of the above mixed precision
CG method further.
In the CG, there are two iterative loops (inner and outer) that are
completely independent of each other.
So information such as the residual vector or directional vector is
not shared between the inner and outer loops except corrected solution
$y$.
Hence, it is possible to improve the performance of CG by transferring
these informations to inner loops as in Refs.~
\cite{Strzodka:2006:PMP:1170135.1170459}
Because informations from the previous run of the inner solver can be
recycled for solving the next problem in the inner loop.
To transfer the information, the initial value of inner loop vectors
are set as
\begin{equation}
\mathbf{y_0} = 0\mbox{,}\quad 
\mathbf{r^{y}_0} = \mathbf{r_k}\mbox{,}\quad 
\mathbf{d^{y}_0} = \mathbf{r} + \beta \mathbf{d} 
\end{equation}
where $\mathbf{r^{y}_0}$ and $\mathbf{d^{y}_0}$ are the initial
residual vector and the initial directional vector in the inner CG
loops, respectively.
The main idea of this initial condition is preserving the orthogonal
descend direction $\mathbf{d^{y}}$ from the previous loop.
As a result, CG method can continue the iteration in the orthogonal
direction, which achieves significantly faster convergence.

The next problem is that using single precision calculation can induce
the accumulation of round-off errors.
This can cause discrepancy between iterated residual and true residual
as the iteration proceed.
Furthermore, the solution vector is also vulnerable to this kind of
error.
To prevent these errors, there are two solutions suggested in the
market: one is the reliable update method and the other is the group
wise update method
\cite{Sleijpen:1996:RUR:226575.226585,Clark20101517}.
The reliable update method is that if the inner-loop residual is
decreased by $\epsilon^{in}$ compared to maximum of all the previous
residuals ($\mathbf{\|r^{y}_n\|} < \epsilon^{in} \,
Max(\|\mathbf{r^{y}}\|)$), it updates the single precision residual 
vector $r^{y}_n$ by the double precision residual 
$\mathbf{r_{k+1}} = \mathbf{b} - A \mathbf{x_{k+1}}$ after
we reconstruct the double precision solution vector 
$\mathbf{x_{k+1}} = \mathbf{x_k} + \mathbf{y_n}$.
By using this method, single precision iteration is automatically
restarted so that accumulated round off error of residual vector is
corrected periodically.
The group wise update method is that if the residual satisfies the
condition
($\mathbf{\|r^{y}_n\|}<\epsilon^{in}Max(\|\mathbf{r^{y}}\|)$), we
obtain a double precision solution vector
($\mathbf{x_{k+1}}=\mathbf{x_k}+\mathbf{y_n}$),
recalculate the double precision residual vector, and then
reset the single precision solution vector $\mathbf{y_{n+1}} = 0$
while preserving the single precision residual vector as
$\mathbf{r^y_{n+1}} = \mathbf{r_{k+1}}$.
By updating this way, we can prevent th irregular summation of round-
off errors in the solution vector.

We apply both of these methods to our CG code.

All of the inner products are calculated in double precision.
After applying these methods, the mixed precision CG runs efficiently
and correctly without any overhead due to round-off errors.

\subsection{Mixed CG implementation on CUDA}
Getting best performance in CUDA programming is a very complicated problem. 
In our previous paper \cite{POS:Lattice2010.230}, several
optimization methods were applied.
Roughly, we apply 4 optimization methods: coalesced memory access,
register and shared memory, data compression (SU(3) reconstruction),
and extra optimization were used in Ref.~\cite{POS:Lattice2010.230}.
Here, we use the mixed precision method in addition, and adjust the
program to accommodate a single precision calculation.

These optimization methods are not the cure-all solutions.
The best performance can be achieved by trial and error, when applying
these optimization methods.
The data size in single precision is half of that in double precision.
In CG, the main bottle-neck lies in the data I/O.
Naively we expect that the performance will be doubled at least using
the mixed precision update.
Using the shared memory has bank conflict problem in double precision
calculation \cite{nvidia_ref}.
However, the shared memory does not have the same problem in single
precision.
Since the single precision calculation is dominant (99.8\%) in the
mixed precision CG, we can use the shared memory as a fast buffer, which
lowers the usage of registers.
The gives an allowance in high CUDA occupancy rate, which enhances 
the performance of CG.
In practice, we store the $\eta$ values and position vector in the
shared memory.
If the number of threads per block is 192, the size of used shared
memory is $192 \times 2 \times 4 \times 4\text{bytes} = 6144
\text{bytes}$.
The register memory is used as a I/O buffer memory.
Because the \emph{NVCC} compiler automatically choose the variables on
the register, it makes the CUDA occupancy low sometimes.
In our case, by restricting the number of used registers lower than 42
and with appropriate usage of shared memory, we can achieve better
CUDA occupancy(=50\%).

We do not use the 8 parameter SU(3) reconstruction
\cite{Clark20101517} for data compression mainly because it causes a
big round-off error.
Hence, we use 10 parameter SU(3) reconstruction method.
Double precision calculation is the same as explained in our previous
paper \cite{POS:Lattice2010.230}.
Nevertheless, the performance of entire double precision sequences are
also improved by adjusting the packet size of data transfer.

\begin{table}[h!]
\begin{center}
\begin{tabular}{c || c | c }
\hline
 & Double Precision(ms) & Single Precision(ms) \\
\hline
GPU processing time & 2.77 & 1.04 \\
boundary data collect & 0.8 & 0.38 \\
memcpy GPU to CPU & 1.3 & 0.83 \\
MPI communication & 2.1 & 1.01 \\
memcpy CPU to GPU & 1.3 & 0.83 \\
\hline
Total time(measured) & ~8.3 & ~ 4.1 \\
\hline
GFLOPS(measured) & 19 GF & 36 GF\\
\hline
\end{tabular}
\end{center}
\caption{A single Dirac operation time table at single and double precision 
  calculation. GFLOPS is derived from entire CG sequence, not from the 1 
  dirac operation.
}
\label{tab:gpuperf}
\end{table}
In table.\ref{tab:gpuperf}, we present the elapsed time of a single
Dirac operator calculation.
The data transfer is dominant: $2/3$(double precision: DP), and
$3/4$(single precision: SP).
The floating point calculation is sub-dominant: $1/3$ (DP) and $1/4$
(SP).
If we drop out the network communication of MPI and PCI-bus, then the
pure GPU performance is 145(SP), and 55(DP) giga flops.
So, in order to optimize the CG code using multi GPUs, we must focus
on the data communication.

By increasing the data packet size as large as possible, we can
achieve the maximum memory bandwidth in the data communication.
If we can overlap cudamemcpy with MPI communication by using
asynchronous communication, then we can reduce the total communication
time by almost 1/3.
But for this, we need to use the \emph{GPU direct 1.0
  technology}\cite{GPU_direct} and this function is only supported on
\emph{Tesla} series of GPU not on the \emph{GeForce} series.
So we could not apply this functionality in our machine yet.

Because single precision FLOPS in table.\ref{tab:gpuperf} is less than
twice of double precision case, it may look a little bit weird.
But for entire program sequence, there are additional process to
launch the CG operation, such as loading data in CPU memory and
rearranging data for the coalesced access.
Because single precision calculation can finish the job faster, entire
performance in single precision is more sensitive to these extra
tasks.
So the actual single precision performance is slightly less than
twice.

\section{Future Perspective}
In the version of CUDA(< 4.0), even if two or more multiple GPUs are
connected to the same computing node, we need memcpy process between
GPU memory and CPU memory to send the GPU data to another GPU memory.
Recently, Nvidia has announced \emph{GPU direct technology 2.0} which
support direct memory copy between different GPUs within a single
machine node \cite{nvidia_ref}.
We plan to implement this new method to our CG code in near future.
However, for off-node communication, we still need memory copy
process between GPU and CPU.
But removing unwanted memory traffic, we can make the whole memory
bandwidth dedicate to the off-node communication only.
In the end, this function will bring performance enhancement on multi
GPU communication.

\section{Conclusion}
By using GPUs, we can get a good performance in the CG algorithm for
staggered fermions.
The final performance is about 145(SP), 55(DP) GFLOPS per GPU(on GTX
480).
We notice that data transfer between GPU and GPU memory is a
main bottle-neck.
For better performance, various optimization methods are used. 
Including the MPI network communication, the performance is
reduced down to 36(SP), 19(DP) GFLOPS per GPU.
The CUDA code of CG with multi GPUs runs in the production mode
to calculate hadron spectrum and weak matrix elements relevant to CP
violation in the neutral kaon system
\cite{ref:wlee-2010-1,ref:wlee-2010-2} at present.

\section{Acknowledgments}
The research of W.~Lee is supported by the Creative Research
Initiatives Program (3348-20090015) of the NRF grant funded by the
Korean government (MEST).

\end{document}